# The Impact of Structural Changes on Learning Capacity in the Fly Olfactory Neural Circuit

Katherine Xie[1,2], Gabriel Koch Ocker[2]

The Pingry School, 131 Martinsville Road, Basking Ridge, NJ 07920[1], Boston University, Boston, MA 02215[2]

# Abstract

The *Drosophila* mushroom body (MB) is known to be involved in olfactory learning and memory; the synaptic plasticity of the Kenyon cell (KC) to mushroom body output neuron (MBON) synapses plays a key role in the learning process. Previous research has focused on projection neuron (PN) to Kenyon cell (KC) connectivity within the MB; we examine how perturbations to the mushroom body circuit structure and changes in connectivity, specifically within the KC to mushroom body output neuron (MBON) neural circuit, affect the MBONs' ability to distinguish between odor classes. We constructed a neural network that incorporates the connectivity between PNs, KCs, and MBONs. To train our model, we generated ten artificial input classes, which represent the projection neuron activity in response to different odors. We collected data on the number of KC-to-MBON connections, MBON error rates, and KC-to-MBON synaptic weights, among other metrics. We observed that MBONs with very few presynaptic KCs consistently performed worse than others in the odor classification task. The developmental types of KCs also played a significant role in each MBON's output. We performed random and targeted KC ablation and observed that ablating developmentally mature KCs had a greater negative impact on MBONs' learning capacity than ablating immature KCs. Random and targeted pruning of KC-MBON synaptic connections yielded results largely consistent with the ablation experiments. To further explore the various types of KCs, we also performed rewiring experiments in the PN to KC circuit. Our study furthers our understanding of olfactory neuroplasticity and provides important clues to understanding learning and memory in general. Understanding how the olfactory circuits process and learn can also have potential applications in artificial intelligence and treatments for neurodegenerative diseases.

# 1 Introduction

Learning is a fundamental cognitive process, and disorders of learning such as post-traumatic stress disorder and amnesia can be devastating. *Drosophila melanogaster* exhibits learned behaviors such as learning sensory associations. The olfactory neural circuit of *Drosophila melanogaster*, also called the mushroom body (MB), is a brain region known to be involved in associative learning, more specifically olfactory learning and memory (Heisenberg, 2003). The MB is a subject of intense research as its structure is relatively simple yet similar across species. The olfactory circuit also closely resembles areas of the brain that are involved in learning and memory, such as the hippocampus and the cerebellum (Campbell & Turner, 2010).

Previous research has focused on the PN-to-KC connectivity within the mushroom body circuit. While the synaptic plasticity of the KC-to-MBON synapses is known to play a crucial role in the olfactory learning process in *Drosophila melanogaster* (Heisenberg, 2003), the exact extent of the role has not been studied extensively. Our study examines how different structural features of the KC-to-MBON circuit contribute to the MB's ability to distinguish between odor classes.

We built a computational model of the MB based on connectivity data collected from an electron microscopic reconstruction of the larval *Drosophila melanogaster* mushroom body (Eichler et al., 2017). Our MB model consists of three main cell types: projection neurons (PN), Kenyon cells (KC), and mushroom body output neurons (MBON) (Fig. 1A). In the MB network, PNs receive input from olfactory receptor neurons, and the Kenyon cells (KC) receive input from PNs and other KCs. MBONs receive input from KCs, modulatory neurons, and other MBONs (Eichler et al., 2017). In our research, we carried out structural perturbations of the KC-to-MBON circuit to quantify the role of various structural features in olfactory learning. The goal is to improve our understanding of the mushroom body, which can advance our knowledge of more complex structures in the brain. This knowledge has the potential to inspire new machine learning techniques in artificial intelligence and treatments for neurodegenerative diseases.

## 1.1 Foundations of the Model

In an effort to model the biological aspects of the *Drosophila melanogaster* mushroom body, we have taken previously observed biological components and implemented them computationally. Specifically, we used a connectivity matrix of the entire MB network from Eichler et al. (2017) and selected neurons on the left hemisphere of the fly brain. The neural network model is composed of

PNs, KCs, and MBONs proceeding in that order (Fig. 1A). We did not include any MBONs that lacked presynaptic KC connections.

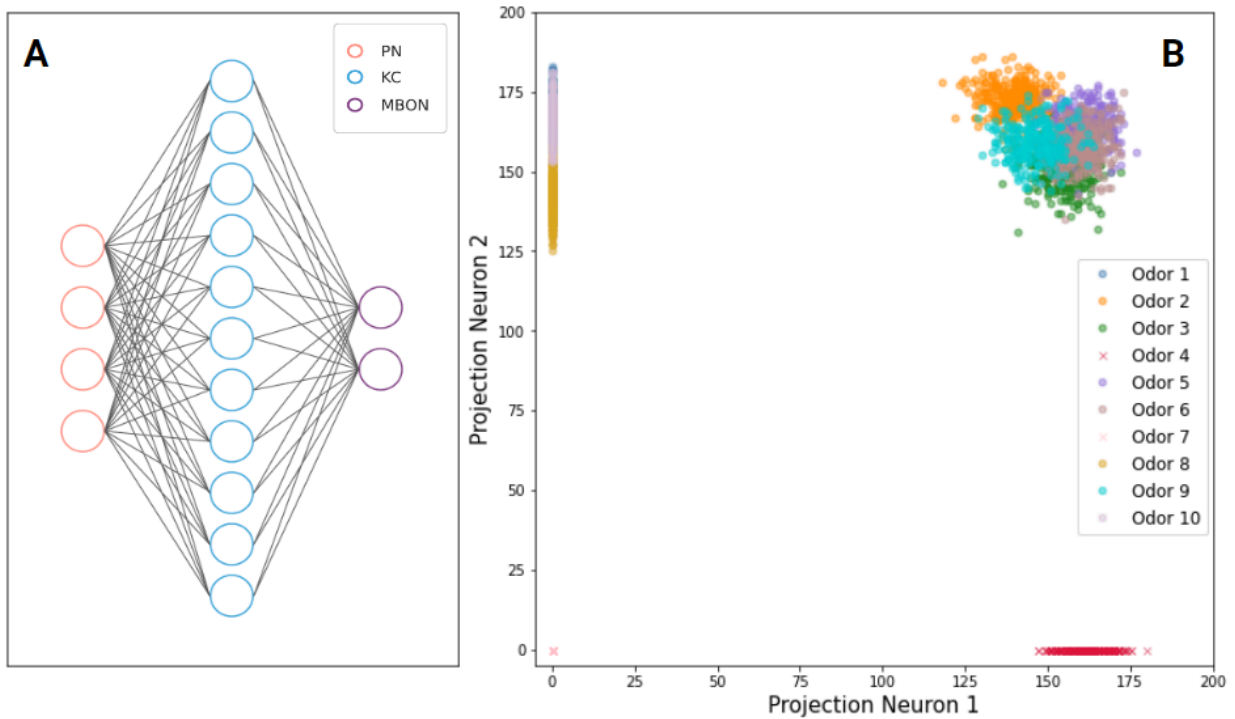

**Figure 1.** Set up for the computational model of the *Drosophila melanogaster* mushroom body. **A)** An approximation of the mushroom body neural network model composed of 40 PNs, 110 KCs, and 18 MBONs and 1 fictional MBON (added for comparison purposes) that takes in inputs from all KCs in the previous layer. **B)** Distribution of PN activity in response to 10 odors. As there are 40 PNs, PN responses are 40-dimensional; however, only two dimensions are displayed here. Distributions marked with "o" indicate that Projection Neuron 2 fires in response to that odor; distributions marked with "x" indicate that Projection Neuron 2 does not fire in response to that odor. Numbers on the axis represent the number of spikes in response to each odor.

Implemented in Python and Jupyter, our model emulates the biological *Drosophila* MB. Different groupings of PNs are found in the glomeruli of the *Drosophila melanogaster* antenna lobe, and each group of PNs typically will only fire in response to certain odors; this typical response is the odor selectivity of PNs (Wilson, 2013). To model this biological response, we created a binomial probability distribution that implements a coin flip model for whether the PN fires or not in response to each odor (Turner et al., 2008). After receiving a stimulus, responsive PNs fire around 162 spikes per second, whereas a KC will likely spike only one or two times within that time frame (Stevens, 2015). If a PN is responsive to a particular odor, then its spike count on a presentation of that odor is sampled from a binomial distribution with 200 trials and the spike probability *r*. To model variability in the PN responsiveness to different odors, we sample *r* from a normal distribution with a mean of 0.8 and a standard deviation of 0.05. We repeat these processes to generate ten separate PN activity distributions, as shown in Figure 1B, in response to the ten odors we use as inputs into our computational model. Each point represents the PN response to a sample of one of the ten odors. Overlaps among the distributions indicate that these odors have similarities in their PN representations.

As shown in Figure 1B, PN 2 does not respond to odor 4 (red "x"'s) and odor 7 (pink "x"'s) as the PN 2 spike value is zero. PN 1 does not respond to these odors: 1, 5, and 8. A cloud of five odors in the upper right of Figure 1B indicates that it would be difficult to distinguish the odors based on only the response activity of PN 1 and PN 2. These odor-evoked PN responses drive KC activity, which in turn provides input to the MBONs (Fig. 1A).

To simulate the olfactory learning process in flies, we trained our neural network model on a 10-odor classification task. We assign each MBON a target response, spike or no spike, in response to each odor. Each MBON must produce the correct output of 1 or 0 in response to its odor-evoked KC inputs. Perceptrons, introduced by Rosenblatt (1958), are used to model each KC and MBON in the MB network. Upon receiving an input point, each perceptron will generate an output based on its weights *wi* and bias θ:

$$y = \sum_i input_i \times w_i + \theta.$$

Mimicking the 5% KC firing rate in response to an odor, we set the KC bias at a threshold value so that each KC would only fire 5% of the time (Honegger et al., 2011). This behavior models the global inhibition KCs receive from the anterior paired lateral neuron (Eichler et al., 2017). At each time step of training, we adjust components of the KC-to-MBON network, including the weights and the bias values, based on the Delta Rule as follows (Widrow & Hoff, 1960):

$$\Delta w_i = error \times input_i \times \alpha,$$

where Δ*w* represents the change in weights and α equals 0.01, the learning rate. The error is the difference between the target and actual output generated by the model. The Delta Rule is applied to the bias values, θ, as well. These adjustments minimize the error in MBON outputs over each time step. Our model was trained for a total of 5,000 time steps. At each time step, the model was given a random input point from any of the ten PN activity distributions. This training process was repeated for 20 realizations.

## 2 Results

Our preliminary analysis consists of various metrics, such as the performance of different MBONs and the evolution of weights. In Figure 2A, most of the MBONs had learned the classification task fairly quickly by around 200 time steps, although there is some variability. Most notably, the learning curves of MBON-o1 (gray), MBON-a2 (orange), and MBON-n1 (magenta) diverge from the rest. MBON-o1 and MBON-a2 take around 400 time steps to learn the task, nearly double the number of time steps of the next slowest MBON. MBON-n1 fails to learn the task, with its error rate (ER) plateauing at around 0.4; this is an instance of a finite training-time effect. Upon further examination, we found that MBON-n1 and MBON-o1 had the two lowest numbers of presynaptic KCs among all the MBONs—6 and 17, respectively. For context, the average number of presynaptic KCs across all the MBONs was approximately 46. MBON-a2 has 63 presynaptic KCs, so the pattern is not perfect. These differences in the competency of various MBONs drive our subsequent experiments and analyses, in which we examine how aspects of the KC-to-MBON network's structure influence these diverse results among MBONs.

There are also noticeable differences in how the weights change over time across different MBONs. Two examples of this are depicted in Figures 2B and 2C. The weights over time of MBON-g1 (Fig. 2C) continue changing for around 120 time steps, whereas the weights of MBON-b1 (Fig. 2B) cease to change after 50 time steps. In addition to the accuracy of different MBONs, another consideration in our experiments is how efficiently each MBON learns the task. Between these two MBONs, MBON-c1 learns faster.

Continuing examinations into the weights of this model, we looked at how weights change after training. We noticed that, for many KCs, there are differences in both the center and the spread between final and initial output weight distributions. For example, consider a 1 claw KC, 2 claw KC, and a 5 claw KC. The final weight distribution of the 1 claw KC has a greater spread than the initial distribution, and the center does not change significantly (Fig. 2D). Figure 2E shows that, after training, the final weight distribution of this 2 claw KC increases in spread and shifts to the left. In other words, the weights' variability increases, and their values decrease overall. The resulting weight distribution of the 2 claw KC's weights differs from the 1 claw KC's as the weight values of the 1 claw KC do not change significantly. The 5 claw KC's final weight distribution (Fig. 2F) does not vary significantly from its initial distribution. The spread increases minimally with only a few weight values shifting negative and some slightly positive.

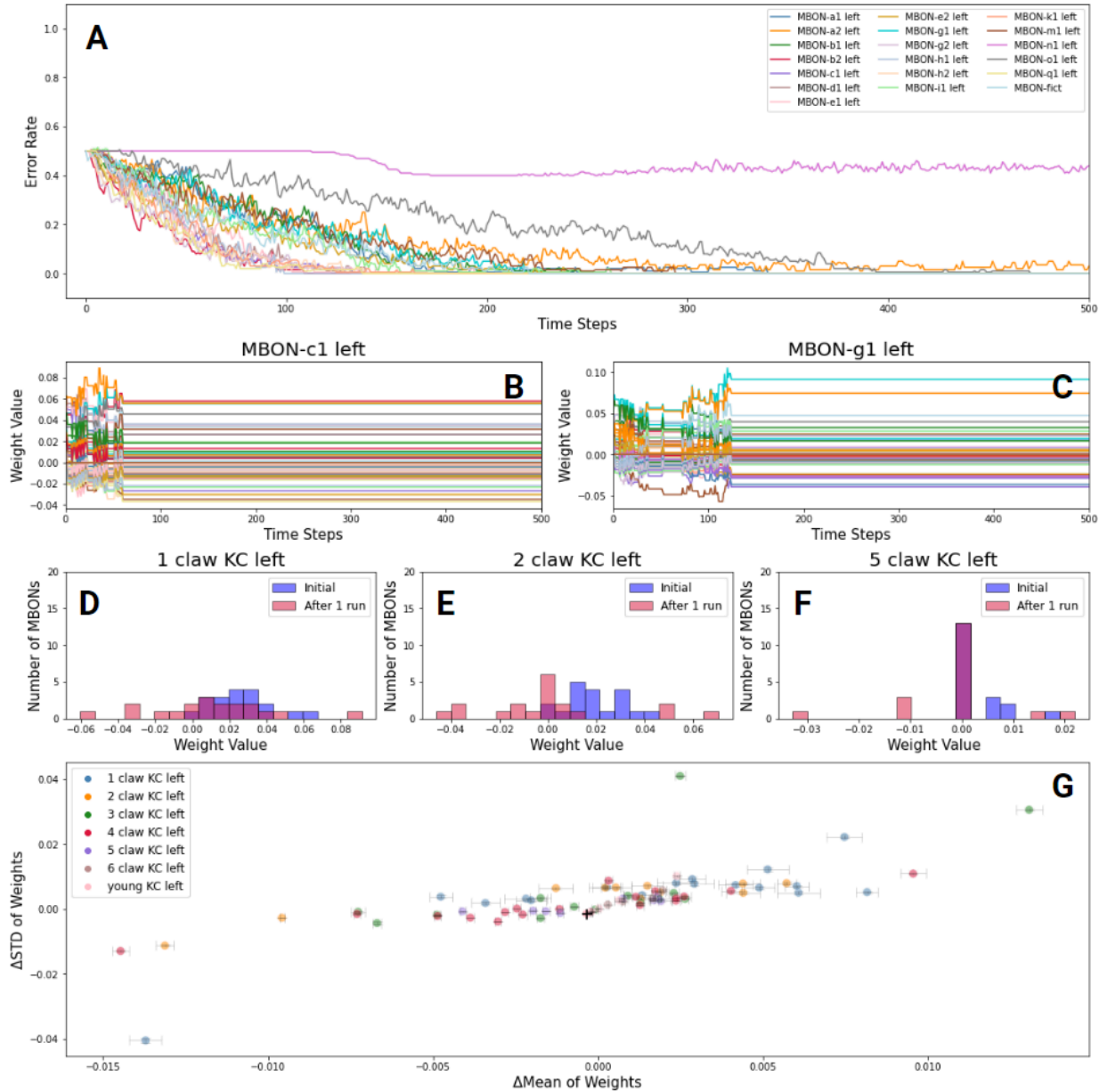

**Figure 2.** Results of training the model and training's effect on components of the model. **A)** Average ER over time across 20 realizations for all 19 MBONs. **B & C)** Weights over time of MBON-c1 and MBON-g1, respectively. Each line represents the changing weight value of a synaptic connection from an MBON's presynaptic KC. **D, E & F)** Initial and final (after 1 realization of training) weight distributions of a 1 claw KC, 2 claw KC, and 5 claw KC, respectively, across all MBONs. The magenta color indicates where the initial and final weight distributions overlap. **G)** Scatter plot of the average across realizations of the change in standard deviation of weights vs. change in mean of weights. Absolute values of the weights are used in the mean and standard deviation calculations. Δ is the change in final weights from the initial set of weights. The seven different KC types are labeled by color.

From the scatter plot in Figure 2G, there is some clustering of 1 claw KCs and 2 claw KCs. The cluster of 1 claw KC points (blue) are distributed around ΔMeans of 0.005 to 0.008; therefore, the average magnitude of 1 claw KC weights is typically stable since that is not a significant weight change. These points have either a slightly positive ΔSTD or no ΔSTD, so the spread of weight values either increases or has no change. Most KCs do not appear to experience any dramatic increase in the spread of their weight values nor a shift in the average magnitude of their weights. There are three outliers: a 1 claw and two 3 claw KCs. The 1 claw KC has the lowest ΔSTD of -0.04, which indicates that its spread decreases. One 3 claw KC has a ΔMean close to zero but a ΔSTD of 0.04, indicating an increase in spread that is greater proportionally than that of the other KCs. The other 3 claw KC has a ΔMean of around 0.013, the greatest ΔMean out of all KCs, which means that the average magnitude of its weights increases more than other KCs. These differing results between KCs raise further questions about the different roles and levels of influence that each KC type has.

In the biological mushroom body, KCs that have fewer claws are typically more developmentally mature (Eichler et al., 2017). These older KCs, we hypothesize, are more involved in the olfactory circuit's ability to learn; younger KCs that have not fully developed are less involved. To test this, we will ablate KCs and measure the impact of this ablation on the model MBONs' ability to learn the odor classification task. As we found that KCs with fewer claws had higher final weights onto the MBONs, we predict that ablating these more mature KCs will affect MBON performance more than ablating those that are younger and have lower weight values.

To examine the differences between KC types, we measured the weight values of the KC output projections as well as the number of projections over time (Fig. 3). During training, the mean weight value of 1 claw KCs remains consistently higher than the other KC types. In contrast, young KCs have the lowest average weight value (Fig. 3A). There is less of a significant difference between the weight values for 2 claw to 6 claw KCs. We observed that the number of output KC projections changes over the course of training as some connections are reduced to a negligible value. With a threshold weight value of 0.005, there were not many significant changes in the number of KC projections over time for each KC type (Fig. 3B). 4 claw KCs and young KCs experienced the greatest decrease in their average number of projections over time. There is a pattern in which the 1 claw KCs have the most output projections, then the 2 claw KCs, 3 claws, and so on. The young

KCs have the least average number of projections, which approaches zero as time passes. From this analysis, we can predict that 1 claw KCs have a considerable influence on MBON performance as 1 claw KCs have, on average, the greatest number of projections to MBONs and the highest weight strengths.

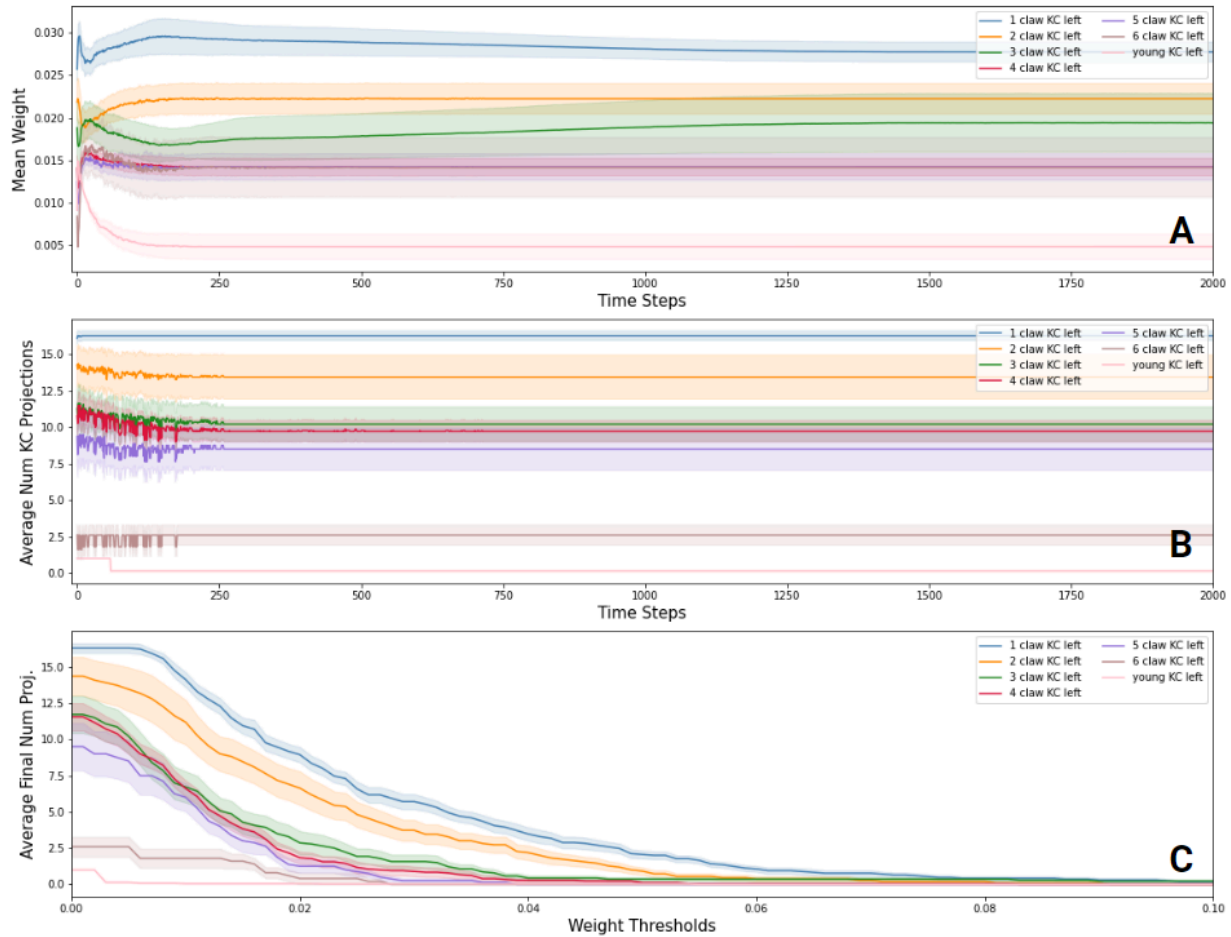

**Figure 3.** Weight statistics of the seven KC types and their output projections. Averages are taken across realizations and then across KCs from each type. Fill between for each line is the standard error across KCs from each type. **A)** Average weight value of each KC type over 2,000 time steps of training. **B)** Average number of KC output projections over time for each KC type with a weight threshold of 0.005. **C)** Post-training average number KC output projections across different weight value thresholds.

After the training process, the average number of projections for each KC type still maintains the previously observed pattern in which 1 claw KCs have the most output projections, then 2 claw KCs, and so on, with young KC having the least number of projections. As the weight threshold value increases, this pattern continues until around 0.06, at which point, most the number of projections for all KC types are close to zero (Fig. 3C). This indicates that most projections have weight values that are less than 0.06.

## 2.1 Ablation Experiments

To further explore the observed differences between individual KCs and MBONs, we performed two types of ablation experiments: random ablation and targeted ablation. When a KC is ablated, we altogether remove that KC from the network by setting all of its synaptic connections to a weight of zero. Biologically, ablation is akin to the process of neuronal apoptosis in the mushroom body. In targeted ablation, KCs are removed from the network based on their total absolute synaptic weight output (Fig. 4B). KCs with greater total weight values will be ablated first; for example, 1 claw KC would be ablated before 5 claw KC and 2 claw KC (Fig. 2D-F). We perform these two experiments to test structural perturbations' effect on the model's performance. We also sought to determine the level of influence that synaptic connection strength has on the model's performance.

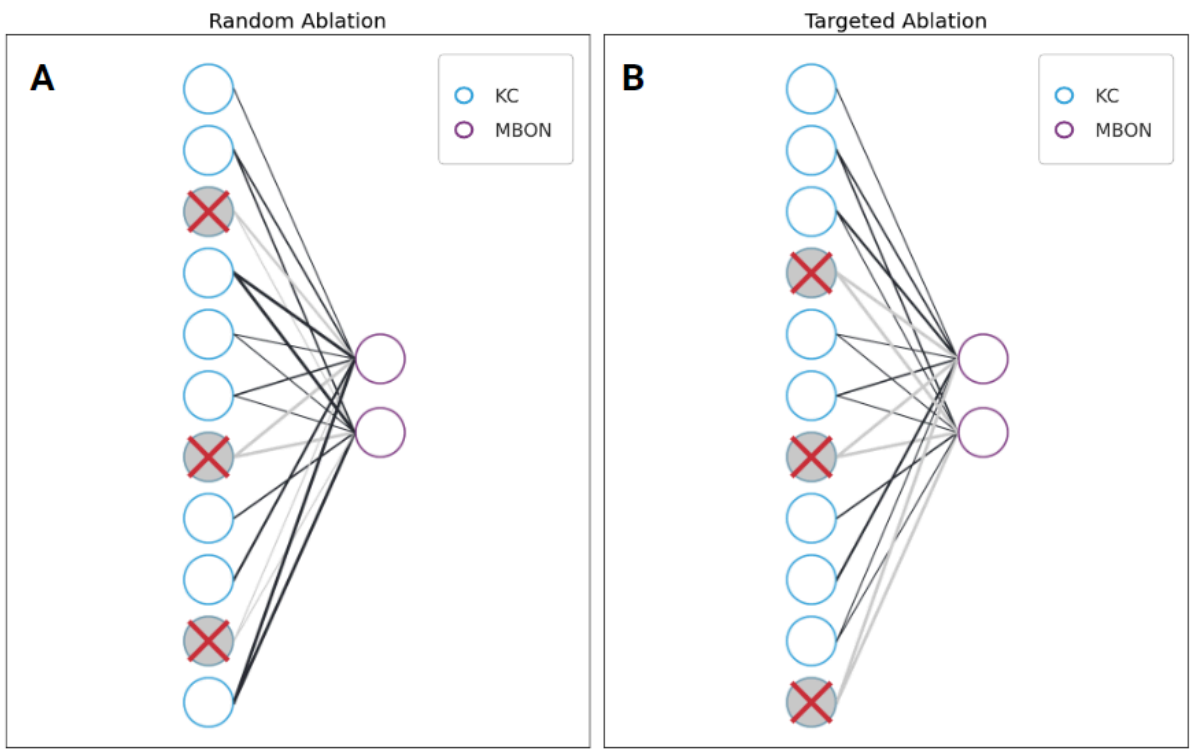

**Figure 4.** Visualization of random and targeted ablation procedures. In each diagram, 3 KCs have been ablated from the network. KCs marked with a red "x" have been ablated. Black lines between nodes indicate that there exists a synaptic connection. Gray lines mark the absence of a synaptic connection. Line thickness indicates the strength or weight value of the synaptic connection between the KC and MBON. **A)** Random ablation of KCs. A certain number of KCs are randomly selected for ablation, meaning all of their postsynaptic connections to MBONs are removed. KCs are ablated regardless of synaptic weight value. **B)**

Targeted ablation of KCs. A certain number of KCs are removed based on their total strength of synaptic connections. KCs with the highest sum of synaptic weight values are ablated first in targeted ablation.

From these two experiments, we collected various metrics and compared the resulting performances of different MBONs after each type of ablation. Figures 4A-C show the ER over 1000 time steps after random ablation of different numbers of KCs. As the number of KCs ablated increases from 1 to 10 to 70, more MBONs take more time steps to learn the task or fail to learn the task. The learning curves of MBON-o1 (gray) and MBON-n1 (magenta) show that these MBONs consistently take longer to learn the task or plateau at a particular ER. Figures 4D-F also show the ER over 1000 time steps after targeted ablation. As the number of ablated KCs increases from 1 to 70, more MBONs fail to learn the task. The MBONs that are quicker to fail are more sensitive to ablation. With 70 KCs removed, almost all MBONs, except for the fictional MBON (Fig. 1A), do not learn at all (Fig. 5F).

It is also observed that, as the number of removed KCs increases, the learning capacity of the MBONs deteriorates, and the rate of the deterioration is quicker in targeted ablation than in random ablation. Figures 5G and 5H confirm this pattern. In the random ablation experiment, most of the MBONs' final ERs did not start to increase until 70 KCs were removed from the model (Fig. 5G). Only MBON-o1 and MBON-n1 do not follow this pattern. MBON-n1's final ER can be attributed to the method we calculated the difference in final ERs by subtracting the control's (KCs removed = 0) final ER of each MBON from the final ERs after each run in which KCs were removed. We discovered that the two MBONs whose ERs increased sooner also had fewer presynaptic KCs; thus, their KCs with presynaptic connections are more likely to be completely ablated.

In contrast to the random ablation experiment, most MBONs' final ERs in the targeted ablation begin to increase at 20 ablated KCs rather than 70. The final ER of most MBONs have plateaued at 60 KCs, meaning they can no longer learn the task. This trend indicates that the total synaptic weight value of each KC is an important factor. Most KCs with high synaptic weight values for each MBON have likely been ablated by that point. Only MBON-fict continues to learn, which makes sense since it still has presynaptic KCs that have not been removed yet (Fig. 5H). Overall, targeted ablation had a more significant impact on the MBONs' ability to learn to distinguish between odors.

Based on the varying weight distributions in Figure 2D-F and the more significant impact of targeted ablation, we conclude that certain KCs play a greater role in allowing the olfactory circuit to learn. We analyzed the differences between KC types and found that during targeted ablation, the KCs with a lower number of claws were often removed first (Fig. 5I). We observed that when a KC is removed, it is a 3 claw KC that is first removed, indicating that 3-claw KC has the highest total synaptic strength. When 15 KCs have been ablated, around 55% of the 1 claw KCs, about 35% of 2 claw KCs, and just under 20% of the 3 claw KCs have been removed. A clear descending trend emerges: at 60 KCs ablated in particular, we observe that 100% of the single claw and 2 claw KCs have been removed, and next 90% of 3, 4, and 5 claw KCs have been ablated. This pattern continues. 0% of the 6 claw KCs and young KCs have been removed as they have lower total synaptic weight outputs. KCs with fewer claws are typically ablated first. Past 60 KCs ablated, these KCs' ablation only increases the final ER minimally, if at all (Fig. 5H), indicating that the KCs with higher numbers of claws have less influence over the MBONs' learning outcome.

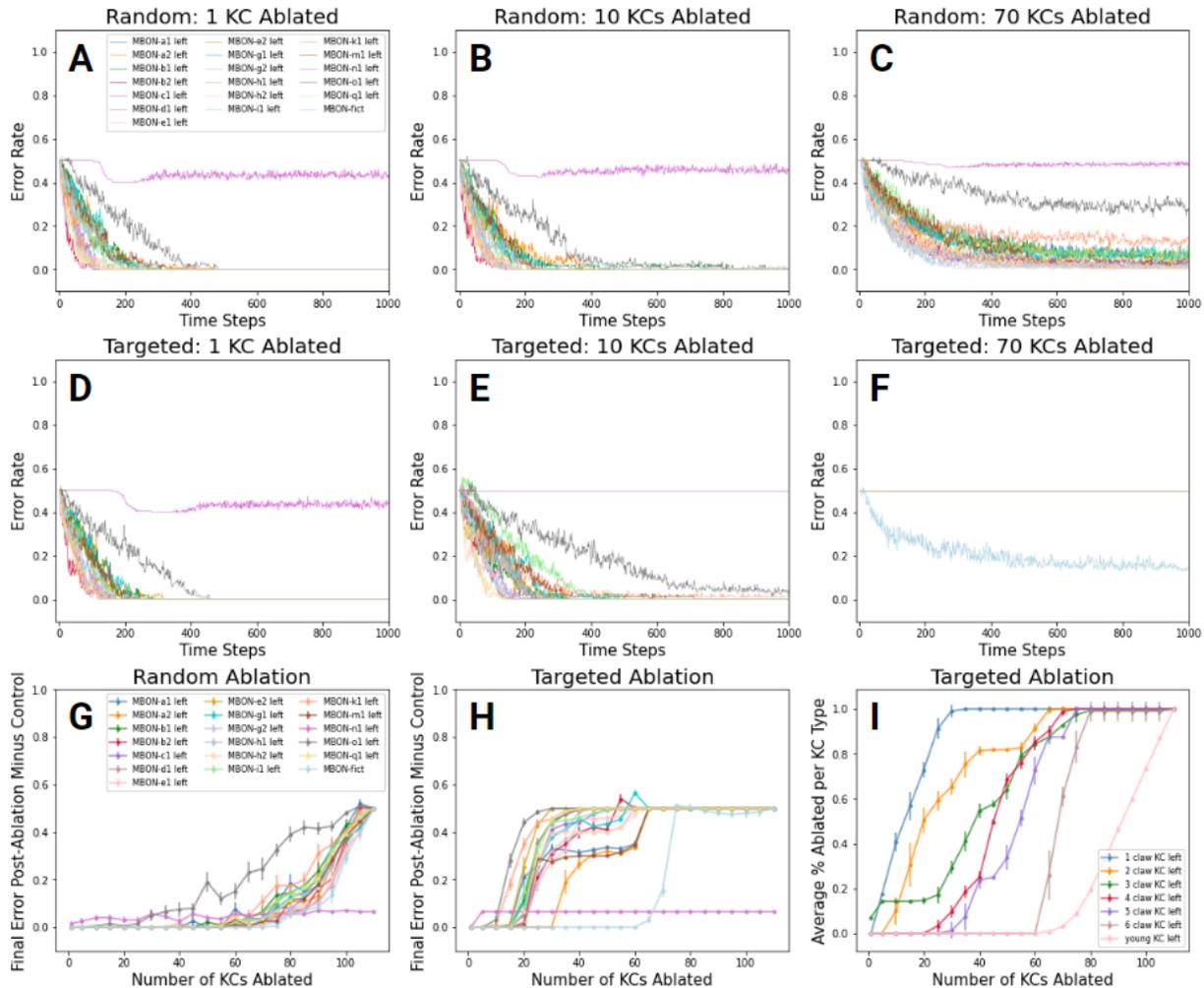

**Figure 5.** Results of the random and targeted ablation experiments. **A, B, C)** ER over time across MBONs for 1 KC, 10 KCs, and 70 KCs, respectively, removed during random ablation. **D, E, F)** ER over time across MBONs for 1 KC, 10 KCs, and 70 KCs, respectively, removed in targeted ablation. **G)** Average final ER after random ablation in relation to the average final ER in the control (0 KCs removed). ERs are averaged over 20 realizations of training. **H)** Average final ER after targeted ablation in relation to the average final ER in the control (0 KCs removed). **I)** Average percentage of KCs ablated that are of each KC type vs. the number of KCs ablated. Error bars in **G**, **H**, and **I** represent the standard error of the mean.

According to Eichler et al. (2017), single-claw KCs are known to be more developmentally mature than other KC types (Eichler 2017). Ablating these single-claw KCs had the most substantial negative impact on the model's learning capabilities (Fig. 5H & 5I). The random and targeted ablation results confirm the hypothesis that KCs with fewer claws are more involved in MBONs' ability to learn the odor classification task. These mature KCs have been integrated into a network structured for associative learning.

## 2.2 Pruning Experiments

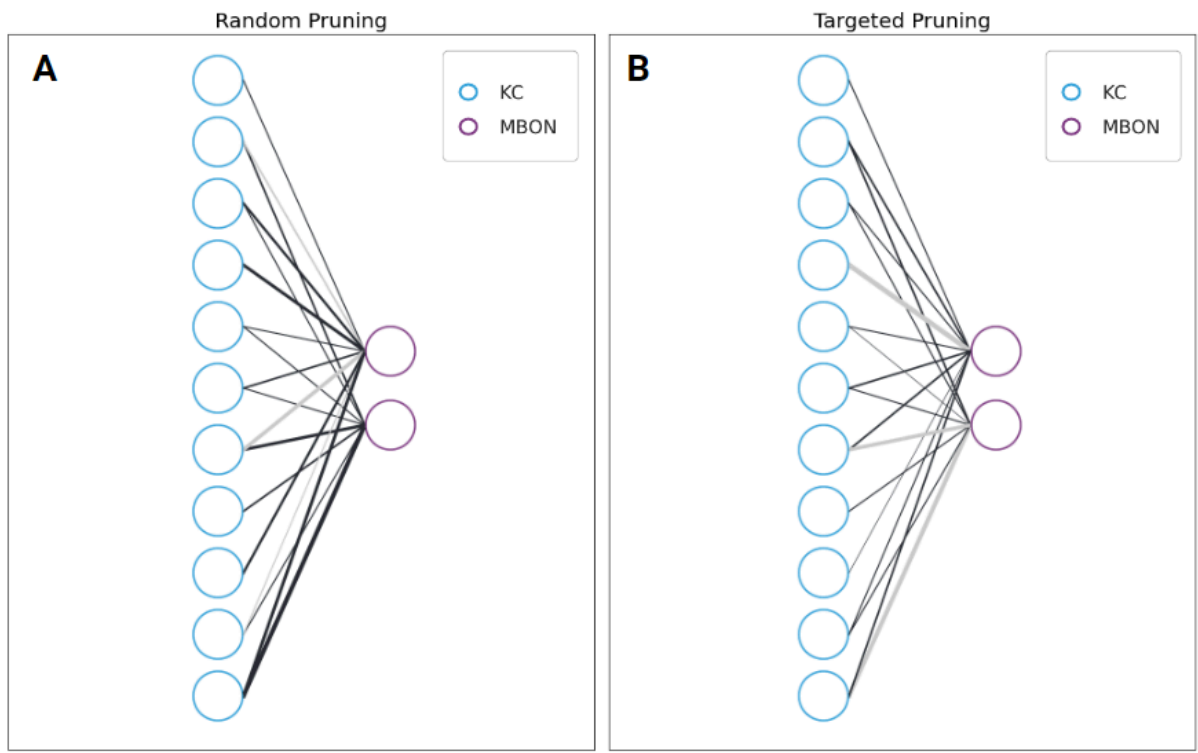

**Figure 6.** Visualization of random and targeted pruning procedures. In each diagram, 3 synapses have been pruned. Black lines between nodes indicate that there exists a synaptic connection. Gray lines mark the absence of a synaptic connection. Line thickness indicates the strength or weight value of the synaptic connection between the KC and MBON. These diagrams detail the pruning of 3 synapses from each MBON. **A)** Random pruning of KC synapses. A certain number of synapses are randomly selected for pruning, and those synapses' weight values are set to zero. KC synapses are pruned regardless of synaptic weight value. **B)**

Targeted pruning of KC synapses. A certain number of KC synapses are removed in the order of which synapses have the highest synaptic weight value.

Building upon the ablation experiments, we conducted KC pruning experiments, in which individual KC synaptic connections are removed. When a KC synapse is pruned, its synaptic weight value is set to zero. We performed both random and targeted pruning, removing one KC synapse, then five to 110 KC synapses in increments of five. In targeted pruning, KC synapses are removed from the network based on their total absolute synaptic weight output (Fig. 6B). For each MBON, synapses with greater weight values will be pruned first. As with the ablation experiments, we performed these pruning experiments to test the effect of structural perturbations on learning. We also aim to compare the effect of pruning and ablation on the MB model's learning performance.

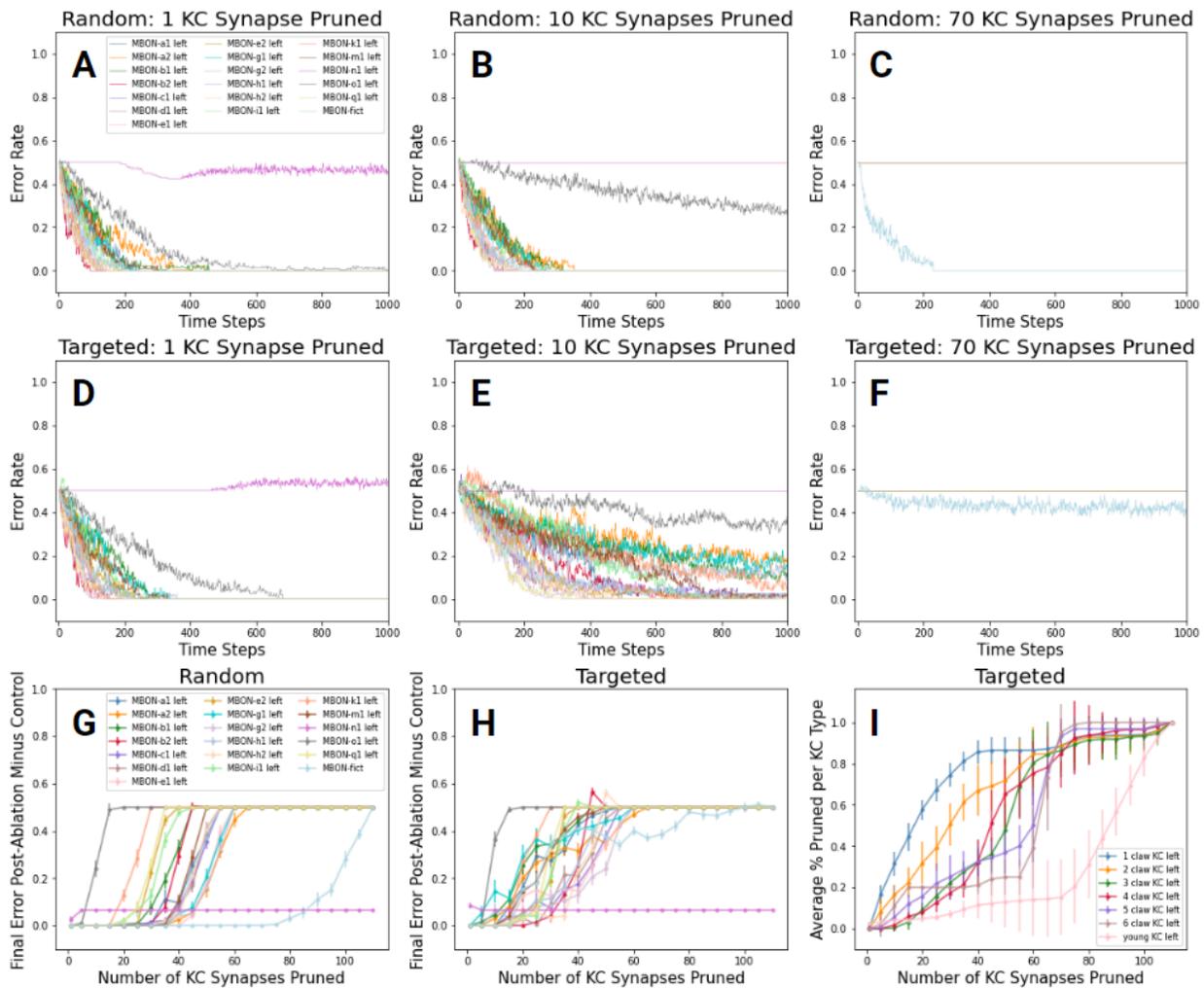

**Figure 7.** Results of the random and targeted pruning experiments. **A, B, C)** ER over time across MBONs for 1, 10, and 70 KC synapses, respectively, removed during random pruning. **D, E, F)** ER over time across MBONs for 1, 10, and 70 KC synapses, respectively, removed by targeted pruning. **G)** Average final ER after

random pruning in relation to the average final ER in the control (0 KCs removed). ERs are averaged over 20 realizations of training. **H)** Average final ER after targeted pruning in relation to the average final ER in the control (0 KCs removed). **I)** Average percentage of KC synapses pruned that are of each KC type vs. the number of KC synapses pruned. Error bars in **G**, **H**, and **I** represent the standard error of the mean.

After collecting various metrics from random and targeted pruning, we observe a similar pattern as we saw with ablation: targeted pruning has a more significant impact than random pruning. Figures 6A-C show the ER over 1000 time steps after random pruning of different numbers of KC synapses from each MBON. As the number of pruned synapses increases from 1 to 10 to 70, most MBONs take more time steps to learn the task or fail to learn the task altogether. Again, the learning curves of MBON-o1 (gray) and MBON-n1 (magenta) show that these MBONs consistently take longer to learn the task or plateau at around an ER of 0.5. Figures 6D-F also show the ER over 1000 time steps after targeted pruning. As with random ablation, as the number of KC synapses pruned increases from 1 to 70, more MBONs fail to learn the task.

Similarly, we also observe that as the number of KC synapses removed increases, MBONs' learning capability diminishes, and the rate of decline is faster in targeted pruning than in random pruning (Fig. 7B & 7E). However, one difference emerges between ablation and pruning; with 70 synapses removed, almost all MBONs, except for the fictional MBON (Fig. 1A), do not learn at all for both random and targeted pruning (Fig. 7C & 7F). This result differs from ablation (Fig. 5C & 5F), in which random ablation of 70 KCs still had MBONs capable of learning. As the number of removed synapses increases, the difference between random and targeted pruning lessens. We explore this phenomenon further in Figure 8.

Figure 7G and 7H illustrate the nuanced difference between the effect of random and targeted pruning on MBONs' ability to distinguish between odors. In Figure 7G, most MBONs' ERs only begin increasing after 20 KC synapses have been removed. For targeted pruning (Fig. 7H), most MBONs' ERs increase when 5 or 10 synapses have been pruned, slightly earlier than in Figure 7G. In both pruning experiments, most MBONs' ERs level off at an ER of 0.5 by the time 60 KC synapses have been pruned. Only MBON-o1, MBON-n1, and the fictional MBON deviate from this pattern. For the fictional MBON, its complete set of KC synaptic connections is likely the cause. As noted earlier for the ablation experiments, MBON-o1 and MBON-n1 have significantly fewer synaptic connections to KCs, so their synapses would all be removed earlier than other MBONs' synapses. Since pruning specifically removes synapses, Figure 7G and Figure 7H confirm our earlier hypothesis that a low number of synapses is linked to their poor performance. Again, note that MBON-n1's low ER can be attributed to our methods of calculating the final ER by subtracting the control ER.

| Kenyon Cell Type | Total Number |
| --- | --- |
| 1 claw | 17 |
| 2 claw | 11 |
| 3 claw | 14 |
| 4 claw | 18 |
| 5 claw | 8 |
| 6 claw | 5 |
| young | 37 |

**Table 1.** The total number of each type of Kenyon cell.

In Figure 7I, we continue to examine the different levels of influence that each KC type plays in olfactory learning through the targeted pruning experiment. 1 claw and 2 claw KC synapses are removed first, which is consistent with the pattern that KCs with lower numbers of claws are removed earlier (Fig. 5I). Interestingly, 4 claw and 6 claw KC synapses are the last ones to be pruned when around 15 synapses have been pruned, and synapses from 3 claw, 5 claw, and young KCs are removed as early as 5 synapses pruned. These results differ from the progression of higher claw KCs being removed at higher percentages first in the targeted ablation experiment (Fig. 5I). This difference can be attributed to the nature of pruning. While targeted ablation removes entire KCs along with all their synaptic connections, in an order based on the total sum of each KCs' synaptic weights, targeted pruning removes specific synaptic connections from each MBON, rather than all the connections of a KC, in an order based on each connection's synaptic weight value. In targeted ablation, young KCs were the last type of KCs to be removed (Fig. 5I); however, in targeted pruning (Fig. 7I), connections with typically lower synaptic weight values—those connected to younger KCs—are removed earlier on. In our computational model, there are 37 young KCs (Table 1). Some MBONs have a majority or only have connections to young KCs, so these young KCs' connections are the only synapses available to be pruned. Since 4 claw and 6 claw KC synapses are ablated later on, these two types of KCs likely have synapses with lower synaptic weight values than those of other KC types.

## 2.3 Pruning and Ablation Comparison

We previously explored the difference between targeted and random ablation and pruning experiments to gauge the importance of synaptic weight value in MBONs' olfactory learning capability. Now, we will compare the effects of pruning and ablation—both have biological origins in the form of synapse removal and apoptosis, respectively.

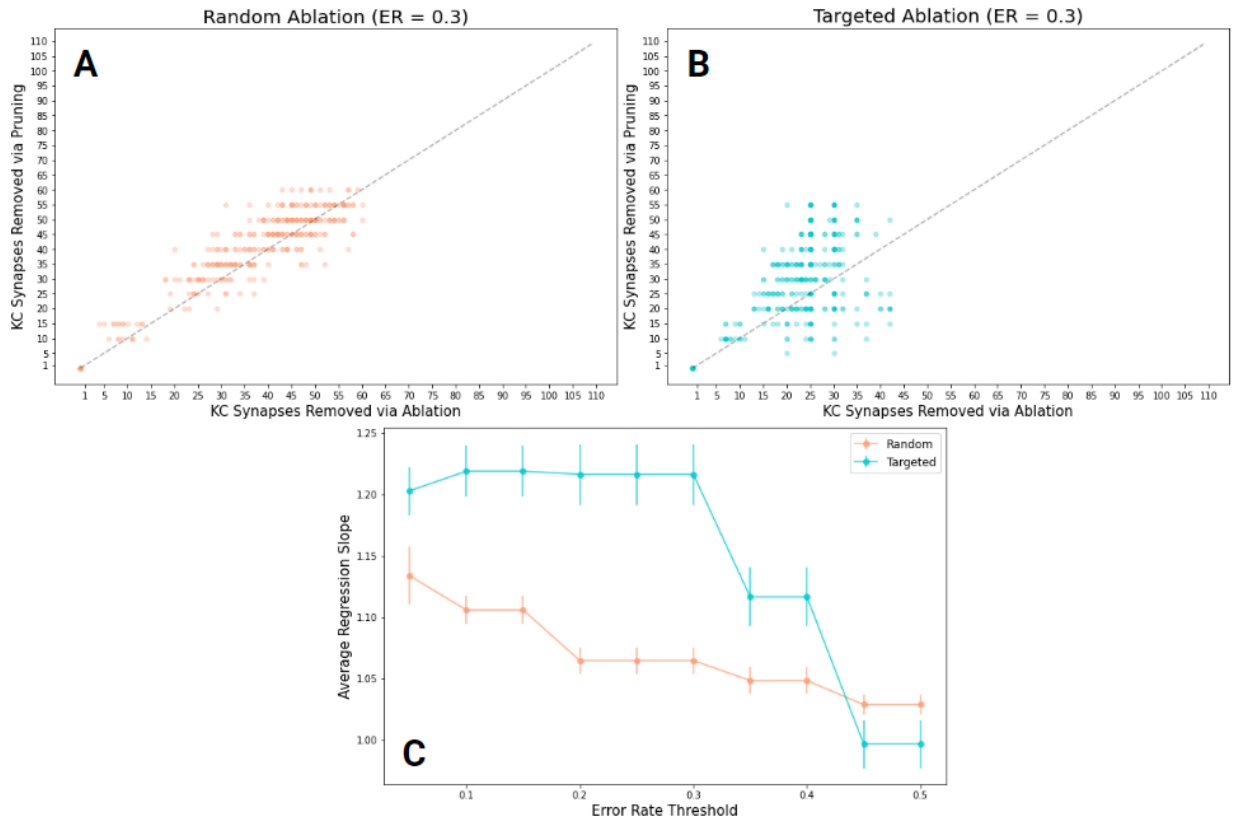

**Figure 8.** Comparison between pruning and ablation's effect on the model's learning capability. Each point represents an MBON, and these plots incorporate data across all 20 realizations of training. The dashed line represents the line y = x. Note: the fictional MBON was removed from these calculations and plots as it would skew the results. **A)** When each MBON reaches an ER of 0.3, the number of KC synapses pruned vs. the number of KCs ablated in random pruning and random ablation, respectively. **A** acts as a control. **B)** Same axes and ER of 0.3 but for targeted pruning and targeted ablation. **C)** Regression slopes for the relationship between KC synapses removed via pruning and those removed via ablation across different ER thresholds from 0 to 0.5.

In Figure 8, we compared the number of KC synapses that need to be removed in pruning vs. ablation for each MBON to reach an ER of 0.3. Each figure has a dashed line with a slope of 1. Figure 8A shows almost an equal number of points above and below the dashed line, indicating that random pruning and random ablation have the same effect on MBONs' learning performance. In contrast, Figure 8B shows more points above the line than below, which means targeted ablation has a more significant detrimental effect on MBONs' learning capacity than targeted pruning has.

We examined the effects of ablation versus pruning across different ER thresholds (Fig. 8C) and discovered that from the ERs in the range of zero to 0.4, targeted ablation has a more significant impact than targeted pruning since the regression slopes are greater than 1. At ER thresholds of 0.45 and 0.5, there is no significant difference between the effects of ablation and pruning as the regression slope is around 1.00. This result indicates that ablation and pruning remove the same number of synapses before the ER reaches 0.45 or 0.5.

## 2.4 PN-KC Rewiring Experiments

In previous experiments, we have examined the KC-to-MBON connectivity. Our results have corroborated our current understanding of KC types and their varying roles in MBON performance. In the PN-KC rewiring experiments, we examine the PN-to-KC connectivity to investigate the fundamental differences between KC types and what factors allow some KCs to have more significant roles than others in determining each MBONs' learning capabilities.

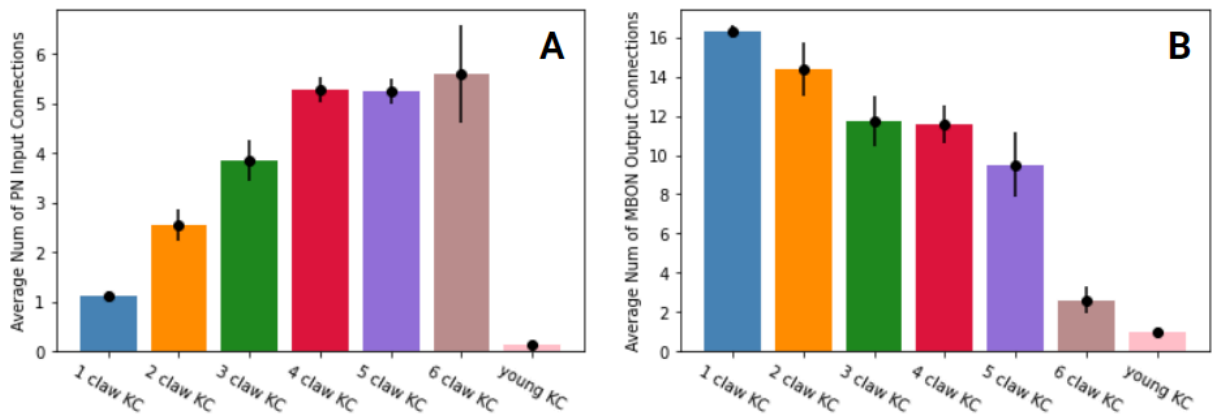

**Figure 9.** Average number of input and output connections for each KC type. **A)** Mean number of presynaptic PN connections to each KC type. **B)** Mean number of output connections for each KC type. Error bars represent standard error of the mean.

On average, each KC has a number of input connections from PNs that is consistent with their KC type—1 claw KCs have one PN input, 2 claw KCs have at least two PN inputs, and so on (Fig. 9A). Interestingly, 1 claw KCs have the greatest number of output connections to MBONs, and young KCs have the least (Fig. 9B). This data supports the conclusion that KCs with fewer claws play a large role in MBONs' performance.

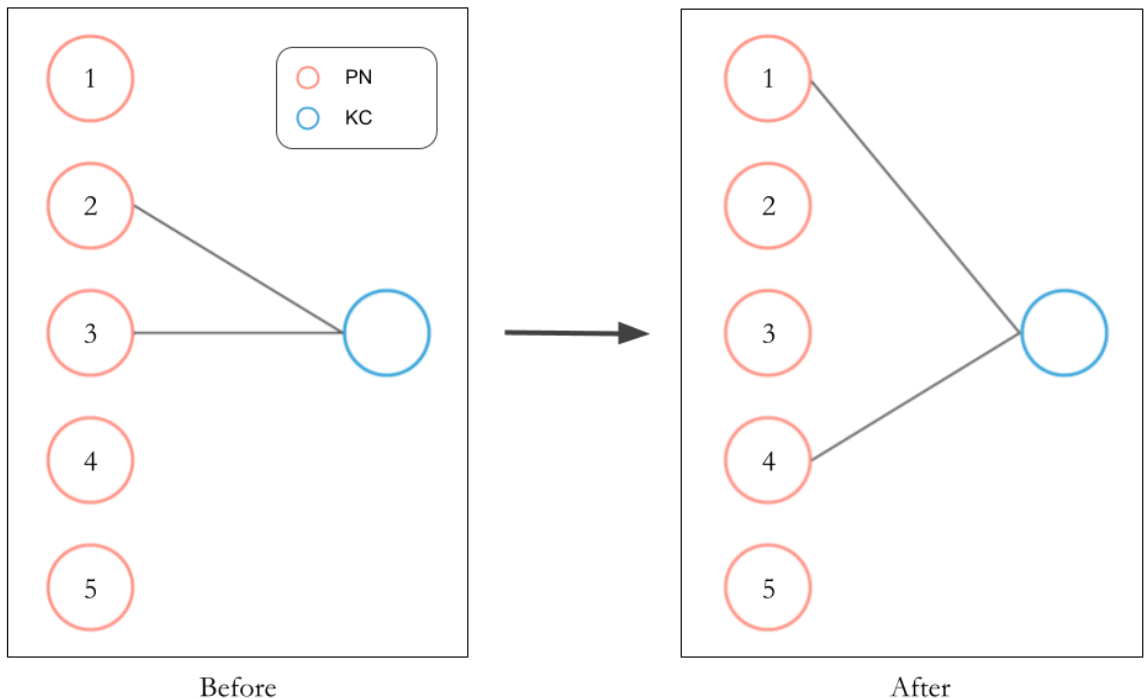

**Figure 10.** Visualization of PN-KC rewiring experiments. The blue circle represents a KC with two presynaptic PN connections from PN 2 and PN 3. After the random rewiring process, the MBON now has two connections from PN 1 and PN 4. The number of presynaptic connections per KC remains constant but the specific presynaptic PNs may change.

Figure 9A shows that the number of claws is a differentiating factor between KC types. To further investigate the relationship between KC identity and its level of influence in the MB network, we performed PN-KC rewiring experiments. In these experiments, we keep each KC's identity but randomize each KC's presynaptic PN connections (Fig. 10). We expected these results to be similar to our first experiment.

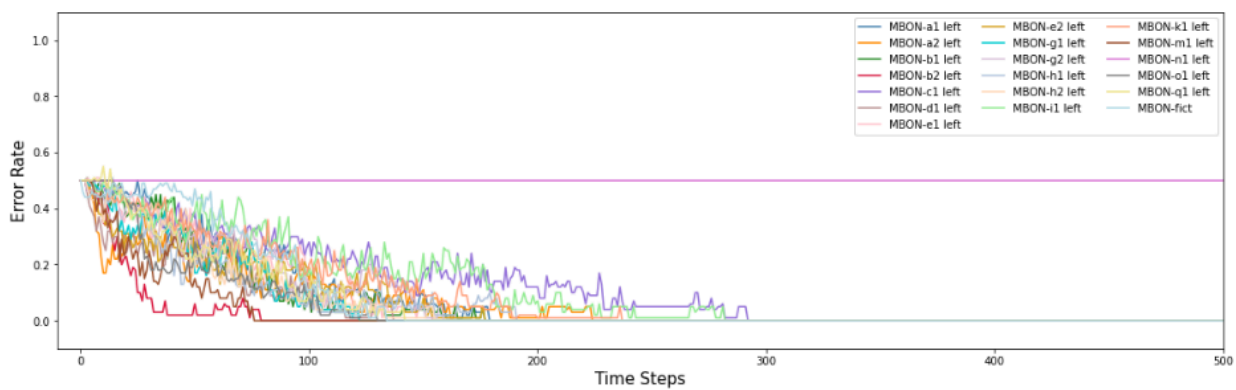

**Figure 11.** Average MBON ER over time with rewiring of PNs and KCs.

The results in Figure 11 resemble those in Figure 2A. MBON-n1 fails to learn the task as its ER remains at 0.5. MBON-o1 is not an outlier, however, in this experiment. Most MBONs achieve an ER of zero in the 100 to 300 time steps range. MBON-b2 and MBON-d1 achieve this around 50 time steps before the majority of MBONs.

After rewiring, MBON-n1 remains an outlier in the ablation experiments, consistent with the results in Figure 2. The random ablation results in Fig. 12A-C are similar to those in Fig. 5A-C, except for MBON-o1 no longer being as much of an outlier. An explanation for this difference could be our rewiring of the presynaptic PNs and KCs of MBON-o1.

In these PN-KC rewiring experiments, we performed targeted ablation based on each KC's total absolute synaptic weight output, as described in Figure 4B. In targeted ablation at 10 KCs ablated (Fig. 12E), MBON-n1, MBON-o1, and MBON-i1 are all outliers. MBON-o1 and MBON-i1 take more time to learn the task than most of the other MBONs. Overall, these targeted ablation results do not differ significantly from our original ablation results (Fig. 5). Figures 15G-I are mainly consistent with the original ablation results as well.

As the identities of KCs are preserved in the rewiring process, we expected the pruning experiment results (Fig. 13) to be consistent with the original pruning results (Fig. 7).

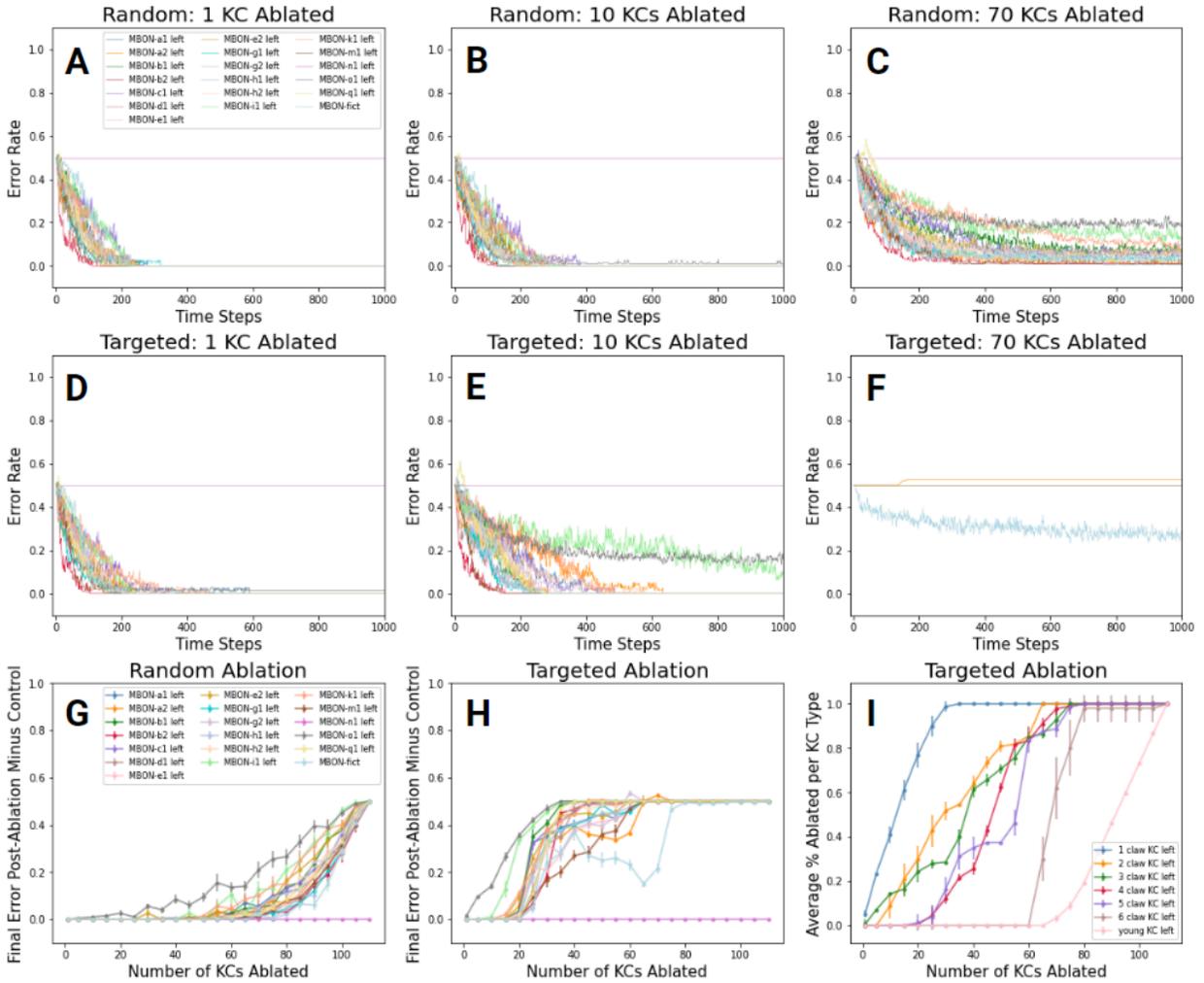

**Figure 12.** Results of the random and targeted ablation experiments with the PN-KC network rewired. **A, B, C)** ER over time across MBONs for 1 KC, 10 KCs, and 70 KCs, respectively, removed during random ablation. **D, E, F)** ER over time across MBONs for 1 KC, 10 KCs, and 70 KCs, respectively, removed in targeted ablation. **G)** Average final ER after random ablation in relation to the average final ER in the control (0 KCs removed). ERs are averaged over 20 realizations of training. **H)** Average final ER after targeted ablation in relation to the average final ER in the control (0 KCs removed). **I)** Average percentage of KCs ablated that are of each KC type vs. the number of KCs ablated. Error bars in **G**, **H**, and **I** represent the standard error of the mean.

The random pruning results (Fig. 13A-C) reflect the original random pruning results (Fig. 7A-C). The targeted pruning results (Fig. 13D-F) also largely reflect the traditional targeted pruning results (Fig. 7D-F). One minor difference is that in Figure 13D, MBON-o1 (gray) does not reach an ER of zero and instead maintains an ER of about 0.1. In Figure 7D, MBON-o1 does achieve an ER of zero within 1,000 time steps.

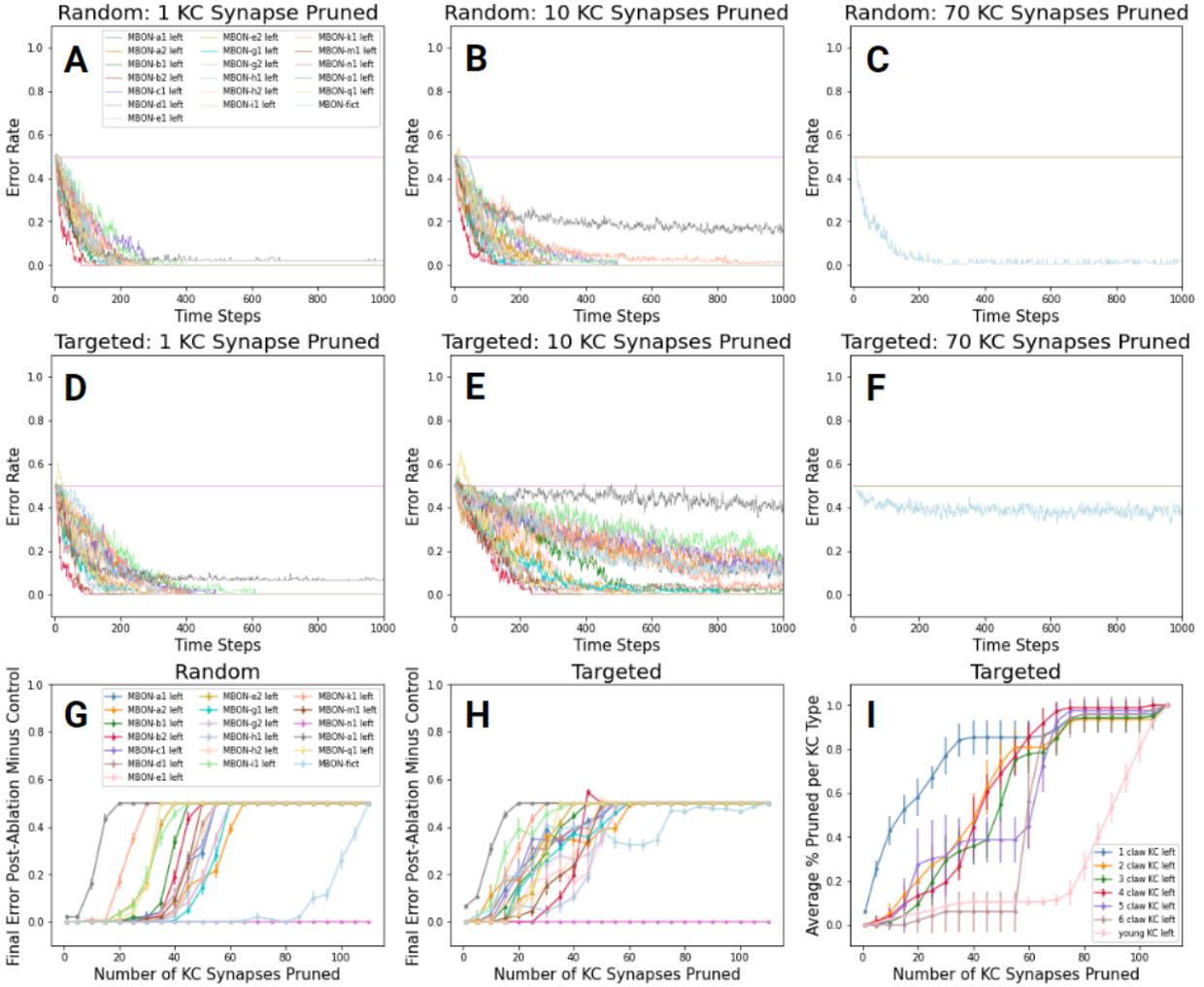

**Figure 13.** Results of the random and targeted pruning experiments with the PN-KC network rewired. **A, B, C)** ER over time across MBONs for 1, 10, and 70 KC synapses, respectively, removed during random pruning. **D, E, F)** ER over time across MBONs for 1, 10, and 70 KC synapses, respectively, removed by targeted pruning. **G)** Average final ER after random pruning in relation to the average final ER in the control (0 KCs removed). ERs are averaged over 20 realizations of training. **H)** Average final ER after targeted pruning in relation to the average final ER in the control (0 KCs removed). **I)** Average percentage of KC synapses pruned that are of each KC type vs. the number of KC synapses pruned. Error bars in **G**, **H**, and **I** represent the standard error of the mean.

The behavior of the average final ER over time for both random (Fig. 13G) and targeted pruning (Fig. 13H) are consistent with the original pruning results (Fig. 7G & 7H). The order of KC synapse type pruned (Fig. 13I) is also largely in accordance with the original pruning results.

Most of the rewiring results were similar to our original control, ablation, and pruning results, with only minor differences. As we maintained KC identities, this outcome emphasizes the importance of

KC type in influencing the performance of MBONs. Next, we explored the consequences of reassigning identities to KCs.

## 2.5 PN-KC Identity Reassignment Experiment

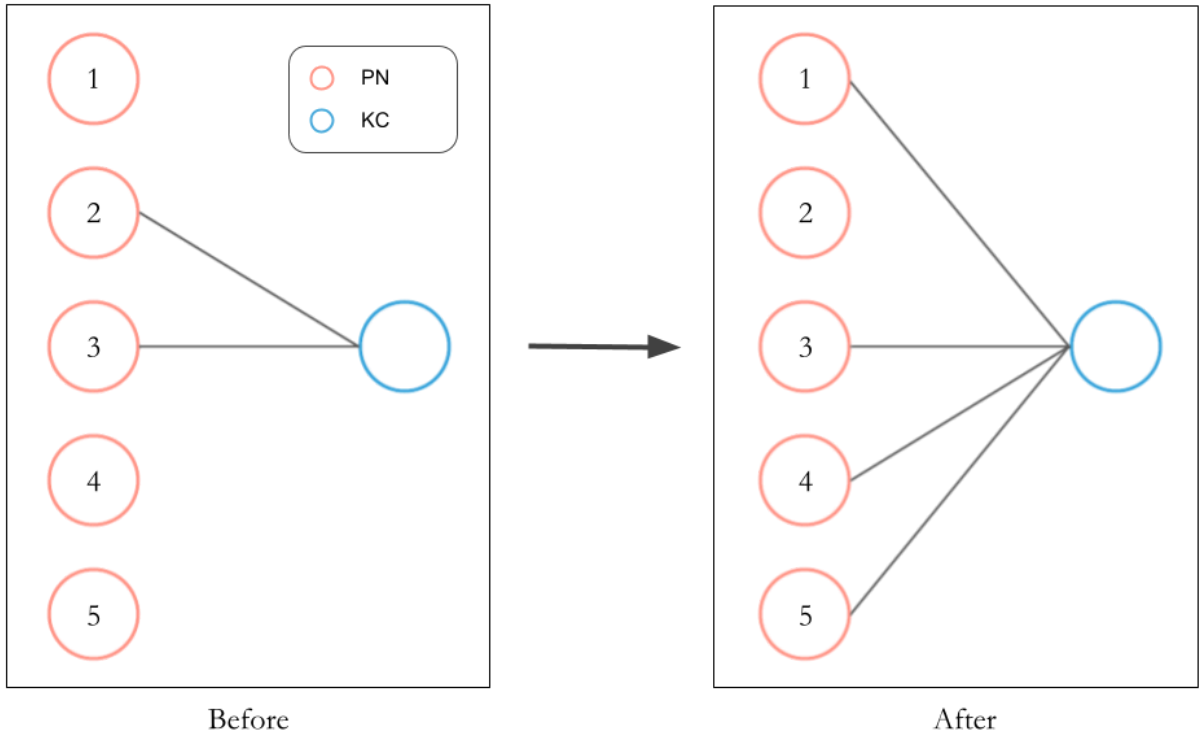

**Figure 14.** Visualization of PN-KC identity reassignment experiment. The blue circle represents a 2 claw KC with two presynaptic PN connections from PN 2 and PN 3. After the random assignment process, this KC's new identity is a 4 claw KC that has four presynaptic PN connections to PNs 1, 3, 4, and 5. Which PNs form synaptic connections with the KC is randomized.

In the previous PN-KC rewiring experiment, we observed that rewiring connections but maintaining KC identities practically had no adverse effect on MBON performance. In the PN-KC identity reassignment experiment, we assigned new identities to each KC by changing the number of presynaptic PNs each KC has (Fig. 14). We also performed the same ablation (Fig. 4) and pruning (Fig. 6) experiments after the reassignment process. Comparing the MBON performance after reassignment to MBON performance in our first experiment will help determine the relative importance of PN-KC connectivity.

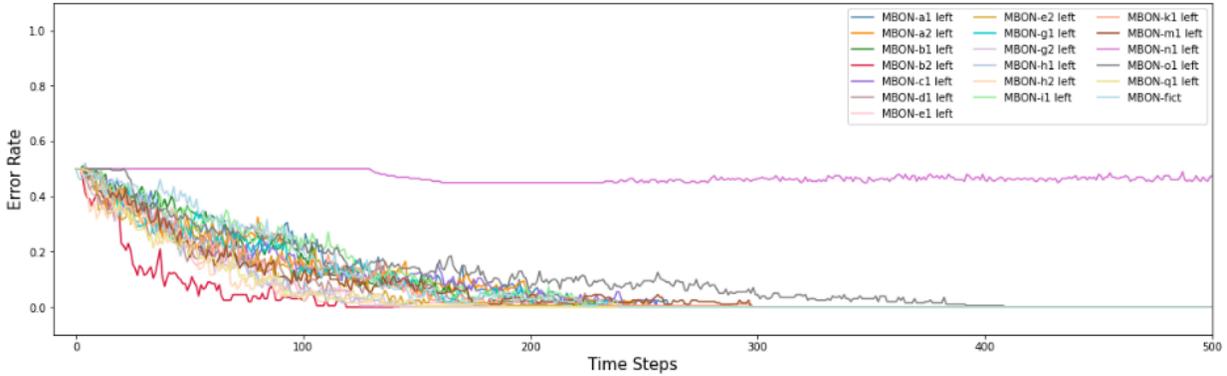

**Figure 15.** Average MBON ER over time with reassignment of KC identities.

After assigning new identities to KCs, most MBONs can learn the task and achieve an error rate of zero except for MBON-n1 (Fig. 15). In the reassignment process, we modified only the number of connections between PNs and KCs while keeping the connections between KCs and MBONs the same. As the network can still achieve low error rates, this result suggests that the number of PN inputs each KC receives is essential to the learning capabilities of the MB network. Figure 9B shows that while the number of claws increases, the number of post-synaptic connections decreases across KC types. Since the reassignment process does not reconfigure these connections, these connections may have less significance. We will further explore the implications of reassignment in learning through ablation and pruning experiments.

The random and targeted ablation experiments differed slightly from our initial ablation results (Fig. 5). In Figures 16A, 16B, and 16D, MBON-i1 takes more time to learn the task than other MBONs. At 70 KCs randomly ablated (Fig. 16C), more MBONs struggled to learn the task. The same was observed after 10 KCs were removed in targeted ablation. These decreases in MBONs' learning efficiency and accuracy indicate that other factors besides the number of PN inputs are crucial to MBON performance.

The final error rate over KCs ablated for random ablation is mostly consistent with Figure 5G. In targeted ablation, the ERs increase more slowly than in Figure 5H. This difference can be explained by the procedure for targeted ablation, which removes KCs based on the sum of their output weight values. Since output weight values are not modified in the reassignment process, the ablation process removes KCs based on their pre-reassignment identities. This procedure also explains why Figure 16I is consistent with Figure 5I.

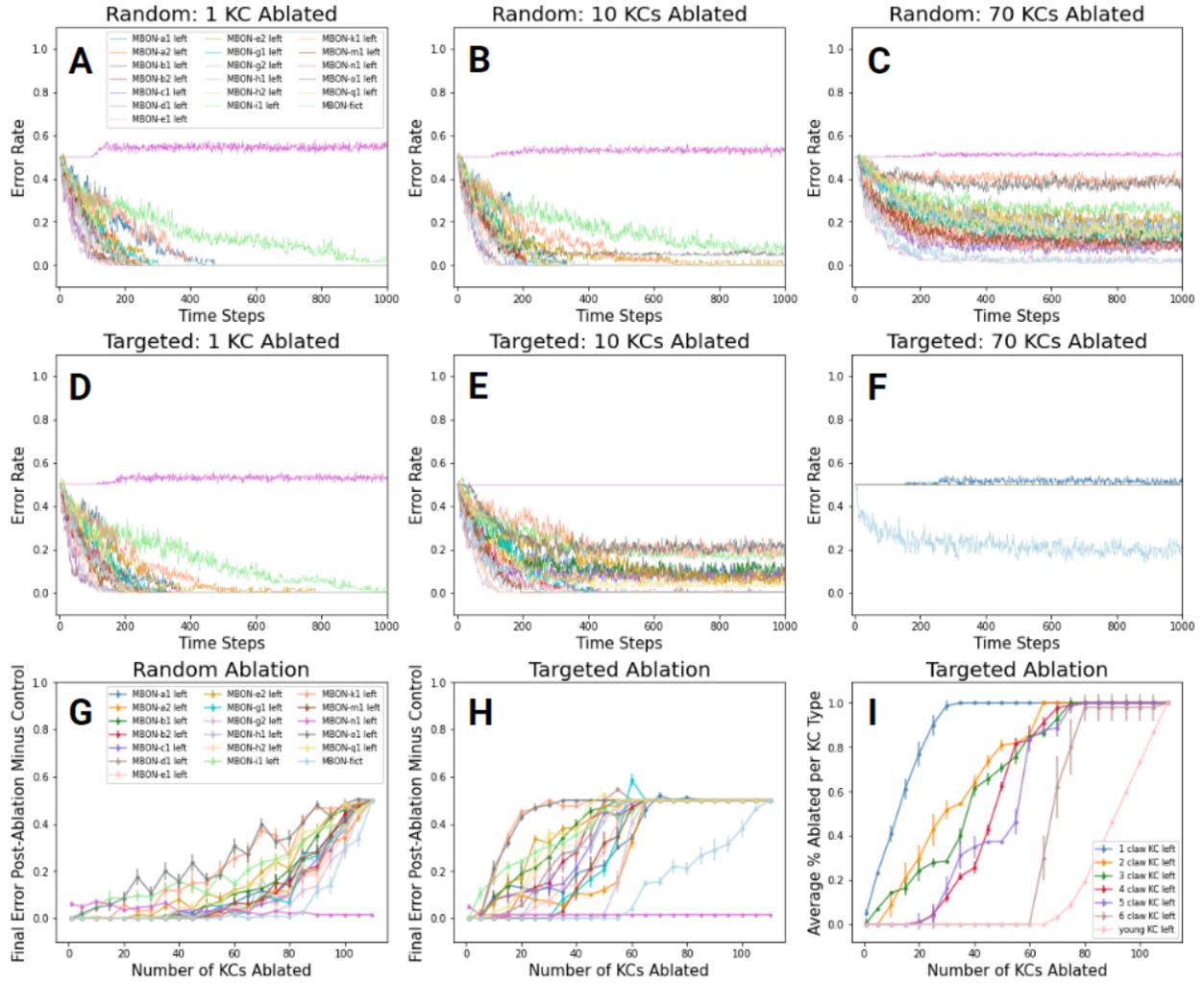

**Figure 16.** Results of the random and targeted ablation experiments with the reassignment of the PN-KC network. **A, B, C)** ER over time across MBONs for 1 KC, 10 KCs, and 70 KCs, respectively, removed during random ablation. **D, E, F)** ER over time across MBONs for 1 KC, 10 KCs, and 70 KCs, respectively, removed in targeted ablation. **G)** Average final ER after random ablation in relation to the average final ER in the control (0 KCs removed). ERs are averaged over 20 realizations of training. **H)** Average final ER after targeted ablation in relation to the average final ER in the control (0 KCs removed). **I)** Average percentage of KCs ablated that are of each KC type vs. the number of KCs ablated. KC types indicated in the legend are based on pre-reassignment KC identities. Error bars in **G**, **H**, and **I** represent the standard error of the mean.

Like the ablation results, many MBONs take significantly more time to learn the task, or their ERs plateau at error rates above zero for both random and targeted pruning (Fig. 17A-F). In random pruning, the final error rate over KC synapses pruned in Figure 17G increases at a similar rate to the final error rate in Figure 7G. In targeted pruning, however, some MBONs' error rate increases at a slower rate (Fig. 17H). The percentage pruned per KC type over the number of KC synapses pruned (Fig. 17I) is consistent with Figure 7I.

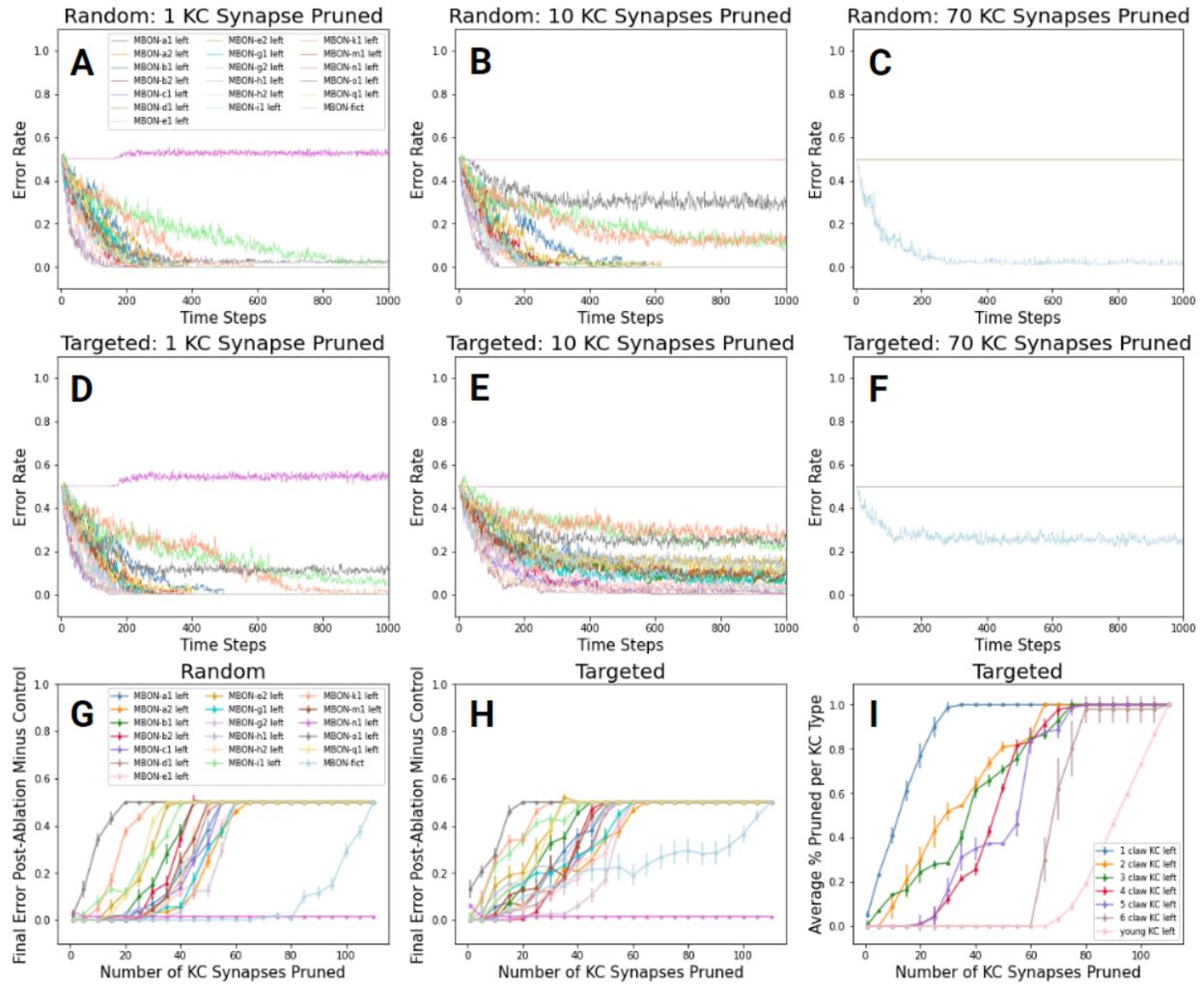

**Figure 17.** Results of the random and targeted pruning experiments with the PN-KC network rewired. **A, B, C)** ER over time across MBONs for 1, 10, and 70 KC synapses, respectively, removed during random pruning. **D, E, F)** ER over time across MBONs for 1, 10, and 70 KC synapses, respectively, removed by targeted pruning. **G)** Average final ER after random pruning in relation to the average final ER in the control (0 KCs removed). ERs are averaged over 20 realizations of training. **H)** Average final ER after targeted pruning in relation to the average final ER in the control (0 KCs removed). **I)** Average percentage of KC synapses pruned that are of each KC type vs. the number of KC synapses pruned. Error bars in **G**, **H**, and **I** represent the standard error of the mean.

While many of the ablation and pruning results were similar to those from our initial ablation and pruning results (Fig. 5 & 7), there were decreases in MBON performance. On the one hand, the similarities highlight the importance of each KC type's specific number of PN inputs. On the other hand, the decreases in learning capabilities emphasize the significant role of KC-to-MBON connectivity.

## 2.6 KC-Constrained Experiments

For all previous experiments, we have constrained the sum of the input weights for each MBON to one. There are two alternatives to this constraint: no constraints or constraining the sum of the KC output weights. To determine the robustness of our previous experiments, we explored the implications of our decision by examining the second alternative.

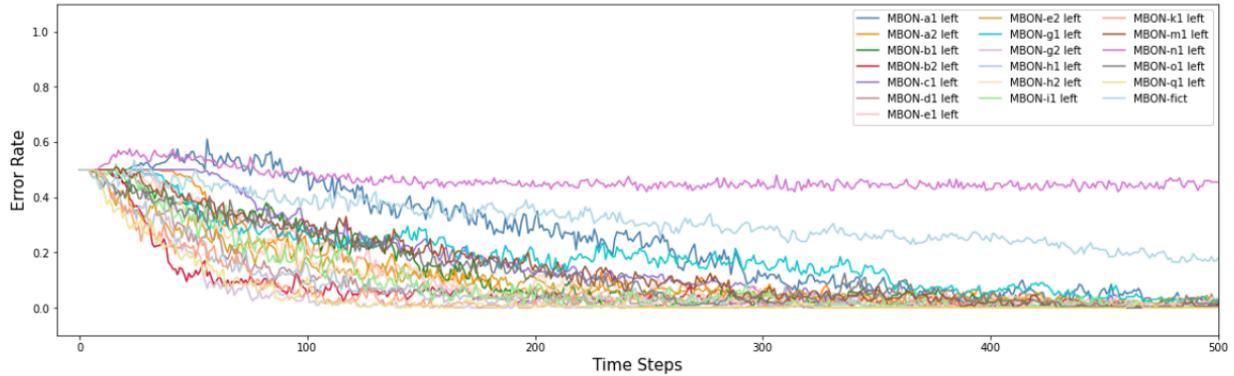

**Figure 18.**   Average ER over time with constraints on the sum of the KC output weights.

In general, the performance of each MBON over time under the KC constraints (Fig. 18) does not differ significantly from the unconstrained results (Fig. 2A). In Figure 18, MBON-n1 remains an outlier as its ER remains around 0.5. However, compared to Figure 2A, MBON-o1 in Figure 18 takes fewer time steps to learn the task. Interestingly, the fictional MBON takes more time steps to learn the task than the other MBONs, excluding MBON-n1. The KC constraints we introduced can likely explain these differences from our previous experiments. We will explore the effect of these KC constraints through KC removal experiments: ablation and pruning.

In the KC-constrained experiments, we observed the same metrics as our previous ablation and pruning experiments and performed the same processes for random ablation, random pruning, and targeted pruning. In targeted ablation, to select which KCs should be ablated first, we normalized the total sum output of each KC by dividing each sum (i.e., one) by the total number of output projections for each KC. KCs with higher normalized sum weight values were ablated first. In other words, KCs with fewer output connections were ablated first.

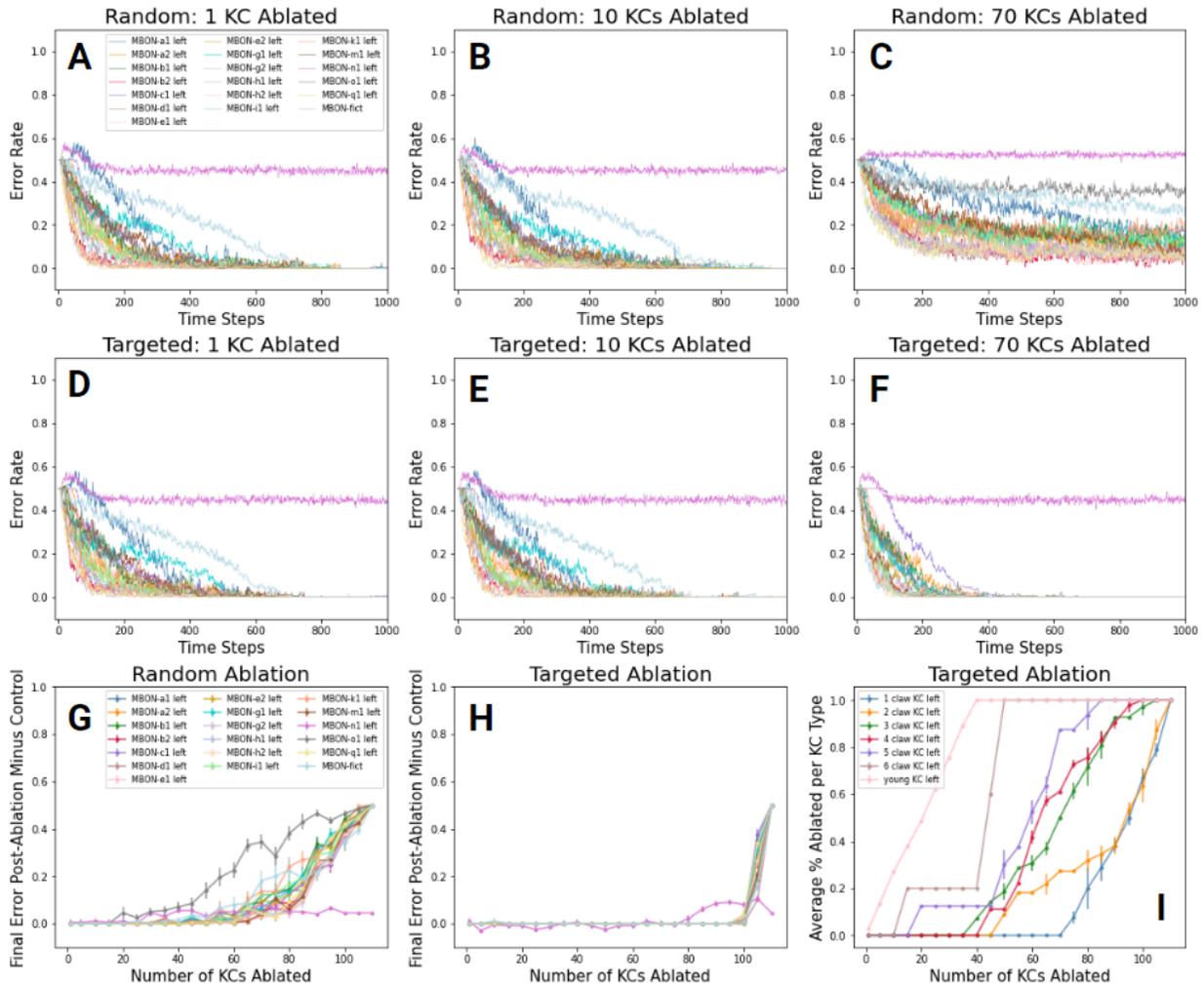

**Figure 19.** Results of the random and targeted ablation experiments with a constraint on the sum of KC output weights. **A, B, C)** ER over time across MBONs for 1 KC, 10 KCs, and 70 KCs, respectively, removed during random ablation. **D, E, F)** ER over time across MBONs for 1 KC, 10 KCs, and 70 KCs, respectively, removed in targeted ablation. **G)** Average final ER after random ablation in relation to the average final ER in the control (0 KCs removed). ERs are averaged over 20 realizations of training. **H)** Average final ER after targeted ablation in relation to the average final ER in the control (0 KCs removed). **I)** Average percentage of KCs ablated that are of each KC type vs. the number of KCs ablated. Error bars in **G**, **H**, and **I** represent the standard error of the mean.

The MBON performance over time (Fig. 19A-F) from the KC-constrained version of the random and targeted ablation experiments contains some differences from the original ablation experiments (Figure 5). One slight difference is that MBON-o1 is no longer as much of an outlier—it takes about the same amount of time as the other MBONs to learn. MBON-n1 remains an outlier as it fails to learn the task. Interestingly, at 70 KCs ablated in targeted ablation, all the MBONs, excluding MBON-n1, took fewer time steps to learn the task than they did at 1 and 10 KCs ablated (Fig. 19F). At 70 KCs ablated, the 6 claw and young KCs have all been ablated, and 2 to 4 claw KCs have been partially ablated. However, none of the 1 claw KCs have been ablated yet (Fig. 19I), suggesting that with only the more mature KCs remaining, the MBONs perform better than with the younger KCs.

Figure 19G does not differ significantly from the original random ablation result (Fig. 5G). In the original targeted ablation experiment (Fig. 5H), the ER of MBONs began increasing as early as 5 KCs ablated. In KC-constrained targeted ablation (Fig. 19H), most of the MBONs' (excluding MBON-n1) ER does not begin increasing until 95 KCs have been ablated. This significant difference can be explained by the type of KC removed. In the KC-constrained version of targeted ablation, none of the more mature KCs, the 1 claw and 2 claw KCs, are removed until 75 KCs have been ablated (Fig. 19I). Only once a majority of the mature KCs have been ablated do the MBONs' ERs increase significantly to around an ER of 0.5. This result reaffirms our findings that mature KCs have a more significant impact on MBON performance.

As young KCs tend to have fewer synapses connected to MBONs, constraining the sum of weights for each KC would mean that the normalized sum weight values for young KCs would be higher than the normalized sum of weights for other KCs with more output connections (i.e., dividing one by a greater number of connections) such as 1 claw KCs. Therefore, in this version of targeted ablation, young KCs are ablated first.

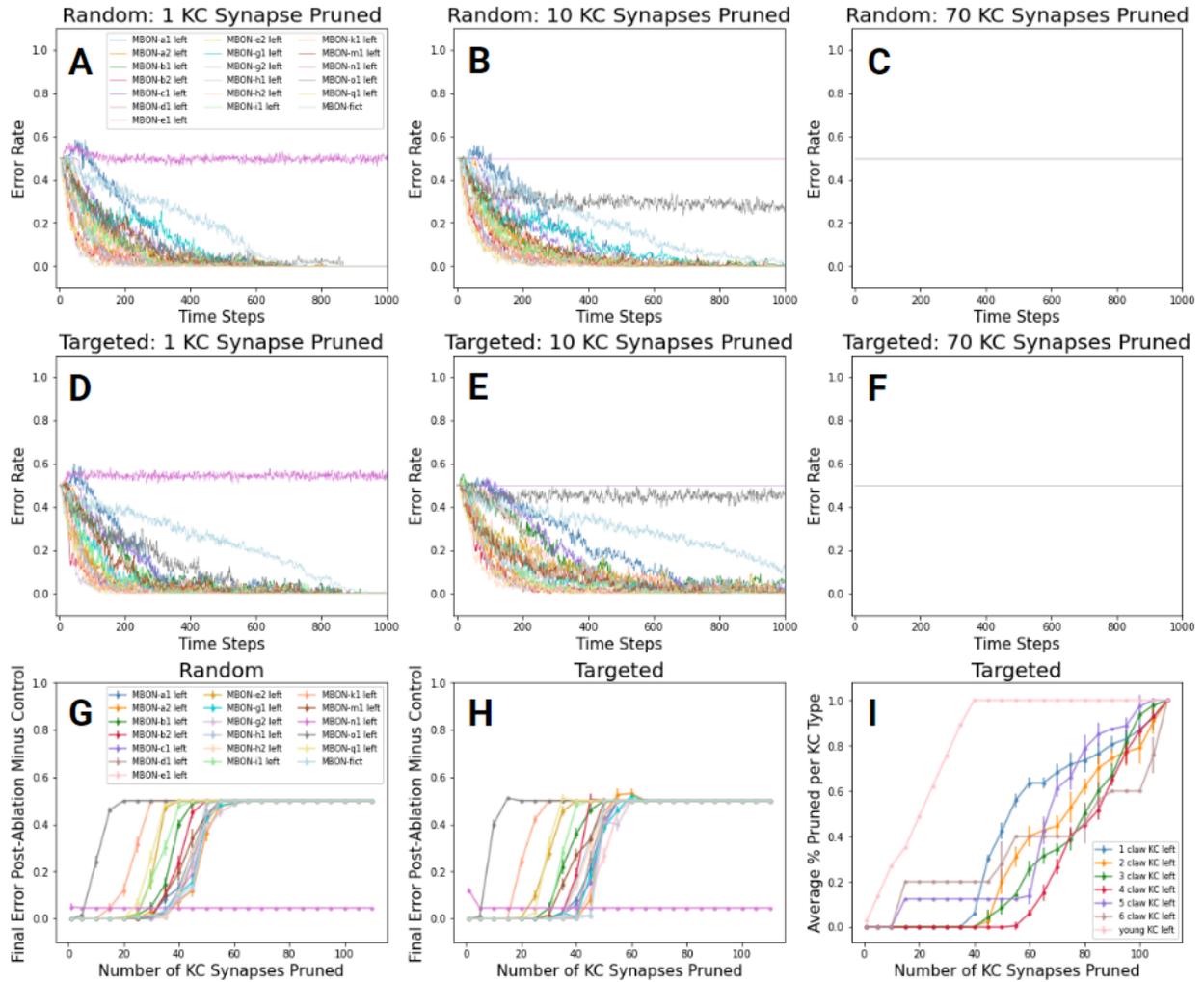

**Figure 20.** Results of the random and targeted pruning experiments with a constraint on the sum of the KC output weights. **A, B, C)** ER over time across MBONs for 1, 10, and 70 KC synapses, respectively, removed during random pruning. **D, E, F)** ER over time across MBONs for 1, 10, and 70 KC synapses, respectively, removed by targeted pruning. **G)** Average final ER after random pruning in relation to the average final ER in the control (0 KCs removed). ERs are averaged over 20 realizations of training. **H)** Average final ER after targeted pruning in relation to the average final ER in the control (0 KCs removed). **I)** Average percentage of KC synapses pruned that are of each KC type vs. the number of KC synapses pruned. Error bars in **G**, **H**, and **I** represent the standard error of the mean.

Compared to the original pruning experiments (Fig. 7), the KC-constrained pruning results (Fig. 20) do not differ significantly. In random pruning (Fig. 20A & 20B), when 1 KC and 10 KC synapses have been pruned, the fictional MBON (MBON-fict) has a slower learning rate than most of the other MBONs, though its error rate reaches 0 at the same time as most of the MBONs. MBON-o1 is not as much of an outlier in Figures 20A and 20D. However, it struggles to learn the task once 10 KC synapses have been pruned. At 70 KC synapses pruned (Fig. 20C & 20F) for both random and targeted pruning, none of the KCs learn the task. This result differs from the original pruning experiment (Fig. 7C & 7F), in which MBON-fict still eventually learned the task.

Examining the relationship between ER and the number of KC synapses pruned (Fig. 20G & 20H), there is not much difference from the original pruning results (Fig. 7G & 6H). In KC-constrained targeted ablation, synapses from young KCs are removed first (Fig. 20I). This result differs from the original pruning experiment (Fig. 7I) as 1 claw and 2 claw KC synapses were pruned first in that case. Figure 20I also has smaller error bars than Figure 7I, meaning there is less variability in which KC types have synapses removed at any point. The ordering of KC synapses pruned—young KCs, 6 claw, and 5 claw KCs, and then later the more mature 1 claw to 4 claw KCs—is consistent with the ordering in KC-constrained targeted ablation (Fig. 19I). Young KCs tend to have fewer synapses connected to MBONs, so constraining the sum of weights for each KC would mean that the weight values for young KC synapses would be higher than the synapses in other KCs with more connections to MBONs such as 1 claw KCs. Therefore, young KCs are pruned first. Eventually, though, 1 claw KCs' synapses begin to be pruned at 35 KC synapses, so the ERs for most of the MBONs begin increasing to 0.5 after that point.

The KC-constrained experiments again emphasize the role of the more mature 1 claw and 2 claw KCs. When young claw KCs were removed first, the MBONs' ability to learn was not impacted and even improved in some cases (Fig. 19F). While MBONs still had connections to mature KCs, they generally succeeded in learning the task.

## 3 Conclusion

In this report, we have detailed the biological basis of our olfactory neural network model and how different structural features impact olfactory training as well as their roles in olfactory learning performance. The results of this study suggest several paths for future research. Building upon the ablation and pruning experiments, we could further look at the removal of specific types of KCs. This ablation experiment would allow us to evaluate further the relative importance of different KC types that we begin to examine in our current research. Another direction would involve conducting experiments on the right side of the MB in addition to our current study of the left side of the MB. Additionally, further adjustments can be made to the computational model based on its biological equivalent, such as introducing recurrent synaptic connections and inhibitory interactions. To conclude, our computational model of the *Drosophila melanogaster* mushroom body circuit can be used in myriad ways to investigate further the roles and functions of structural changes and neural plasticity, in general, of the neural network on olfactory learning and memory. Computational models can prove an important and versatile tool for studying the neural basis of cognitive function.

In a broader context, our findings on the olfactory circuit's connectivity can aid our understanding of the inner workings of neuroplasticity. The mechanisms of the olfactory system and its plasticity are a crucial area of study as the loss or reduction of olfactory senses is often one of the first symptoms of neurodegenerative diseases such as Alzheimer's and Parkinson's. The link between one's sense of smell and memory is also critical (Aguilar martínez, 2018). Other researchers and companies have looked into olfactory training as a method to help patients recover their olfactory sense and slow the progression of neurodegenerative diseases. Kollndorfer et al. (2014) found that olfactory training actually improved neuroplasticity in these patients. Thus, our understanding of neuroplasticity through the structure of olfactory systems can have far-reaching clinical implications.

In the field of artificial intelligence, the bulk of current machine learning algorithms were based on the structure of the visual cortex. However, scientists are now also turning to the olfactory system for fresh inspiration; Its structure is similar to other more complex brain regions, such as the hippocampus, which is implicated in learning and memory. The nature of odors themselves is random and variable, so understanding how the olfactory circuit processes this information can lead to new machine learning techniques based on olfaction. Research has also proven that the inherent generalization ability of the olfactory learning systems is far superior to that of the visual cortex (Huerta & Nowotny, 2009). The olfactory circuit has the potential to adapt to different forms of learning and could provide insights into how more complex cognitive features of the brain can be integrated into AI.

Github: https://github.com/kxie2022/mushroom-body-research